\documentclass[pre,showpacs,showkeys,preprintnumbers,amsmath,amssymb,superscriptaddress,twocolumn]{revtex4}
\usepackage{graphicx}
\usepackage{amsfonts}
\usepackage{amsmath}

\def\u{{\bf u}}

\def\U{{\bf U}}
\def\C{{\bf C}}
\def\M{{\bf M}}

\def\R{{\mathbb R}}
\def\I{{\bf I}}

\def\x{{\bf x}}
\def\tr{{\rm tr}}

\def\T{{\dagger}}

\def\ve{\varepsilon}

\begin{document}

\title{Optimal and sub-optimal quadratic forms for non-centered Gaussian processes}

\author{Denis~S.~Grebenkov}
 \email{denis.grebenkov@polytechnique.edu}
\affiliation{
Laboratoire de Physique de la Mati\`{e}re Condens\'{e}e (UMR 7643), \\ 
CNRS -- Ecole Polytechnique, 91128 Palaiseau, France}

\date{\today}

\begin{abstract}
Individual random trajectories of stochastic processes are often
analyzed by using quadratic forms such as time averaged (TA) mean
square displacement (MSD) or velocity auto-correlation function
(VACF).  The appropriate quadratic form is expected to have a narrow
probability distribution in order to reduce statistical uncertainty of
a single measurement.  We consider the problem of finding the optimal
quadratic form that minimizes a chosen cumulant moment (e.g., the
variance) of the probability distribution, under the constraint of
fixed mean value.  For discrete non-centered Gaussian processes, we
construct the optimal quadratic form by using the spectral
representation of the cumulant moments.  Moreover, we obtain a simple
explicit formula for the smallest achievable cumulant moment that may
serve as a quality benchmark for other quadratic forms.  We illustrate
the optimality issues by comparing the optimal variance with the
variances of the TA MSD and TA VACF of fractional Brownian motion
superimposed with a constant drift and independent Gaussian noise.
\end{abstract}



\maketitle

\section{Introduction}

The statistical analysis and reliable interpretation of stochastic
processes have become indispensable tools in fields as different as
non-equilibrium statistical physics, biophysics, geophysics, ecology
and finances.  Examples range from random trajectories of individual
tracers in living cells
\cite{Saxton97b,Tolic04,Golding06,Arcizet08,Wirtz09,Metzler09,Jeon11,Bertseva12}
to market stock prices \cite{Bouchaud}.  The acquired trajectories are
often unique, either due to the challenges in reconducting an
experiment or reproducing the identical experimental conditions (e.g.,
in living cells), or due to the intrinsic uniqueness of the phenomenon
(e.g., stock prices).  In both cases, a single realization of the
stochastic process has to be analyzed.  Although the problem of
optimal inferences has been thoroughly studied in statistics for a
long time, none of various statistical tools is known to be
universally the ``best''.  For instance, the maximum likelihood
estimators are known to be (nearly) optimal but their implementation
may be too time-consuming or impractical under certain circumstances.
In turn, a much simpler tool of the time averaged (TA) mean square
displacement (MSD) which is broadly used by experimentalists, may be
biased or strongly non-optimal.  The presence of localization errors,
blurring, electronic noises and other acquisition artifacts may
strongly alter the inferred parameters
\cite{Berglund10,Michalet10,Michalet12}.  As a consequence, the search
for optimal inferences is still active, even for simple and well
studied processes such as, e.g., Brownian motion
\cite{Voisinne10,Grebenkov11b,Boyer12a,Boyer12b}.

In this paper, we consider a discrete Gaussian process of $N$ steps,
i.e., a Gaussian vector $\x = (x_1,...,x_N)^\T \in \R^N$, which is
determined by given mean vector $\x^0$ and covariance matrix $\C$.  In
general, the mean vector and the covariance matrix are not known and
have to be inferred from random realizations of the process.  Such an
inference of $N + N(N-1)/2$ unknowns is obviously impossible from a
single realization of $N$ random points $x_j$.  In many cases,
however, the structure of the mean vector and/or the covariance matrix
is expected.  For instance, one-dimensional discrete Brownian motion
(or off-lattice random walk) with a constant drift is defined by
setting $\x^0 = (a+\mu,a+2\mu,...,a+N\mu)$ and $\C_{j,k} = \sigma^2
\min\{j,k\}$, where $a$ is the starting point, $\mu$ is the drift over
one step (i.e., $\mu = v\delta$ where $\delta$ is the step duration
and $v$ the velocity), and $\sigma^2$ is the one step variance (which
is related to the diffusion coefficient $D$ as $\sigma^2 = 2D\delta$).
Choosing a particular class of processes (i.e., choosing the structure
for $\x^0$ and $\C$), one significantly reduces the number of
unknowns, making the inference from a single realization tractable.
For instance, only three parameters $a$, $\mu$ and $\sigma$ have to be
inferred in the above example.

Many standard estimators employed for the analysis of single-particle
trajectories operate with quadratic forms, $\chi = \frac12(\x^\T \M
\x)$, defined by a convenient symmetric matrix $\M$.  Examples are
TA MSD, TA VACF, power spectral density, squared root mean square
displacements, etc. \cite{Grebenkov11a,Grebenkov11b}.  Why different
quadratic forms have been employed?  How can one choose between them?
What is the ``best'' quadratic form to infer the parameters of a known
stochastic process?  The answers to these questions strongly depend on
the studied process and on the chosen optimality criterion.

Inspired by these questions, we consider here a more specific problem
of finding the ``optimal'' symmetric matrix $\M_{\rm opt}$ that would
minimize the variance $\kappa_2$ (or another cumulant moment
$\kappa_m$) of the quadratic form $\chi$, under the constraint for the
mean value $\kappa_1$ of $\chi$ to be fixed.  In \cite{Grebenkov11b},
we briefly mentioned this problem and showed that the optimal matrix
$\M_{\rm opt}$ for discrete centered Gaussian processes (i.e., for
$\x^0 = 0$) is proportional to the inverse of the covariance matrix
$\C$: $\M_{\rm opt} = \lambda\C^{-1}$, with $\lambda = 2\kappa_1/N$.
In this case, the quadratic form $\chi$ has a Gamma distribution:
\begin{equation}
\label{eq:Gamma}
p(z) = \frac{z^{N/2-1} ~ e^{-z/\lambda}}{\Gamma(N/2) \lambda^{N/2}} .
\end{equation}
For instance, the optimal quadratic form for discrete Brownian motion
corresponds to the TA MSD with the unit time lag: 
\begin{equation}
\label{eq:TAMSD0}
\frac12 (\x^\T \M_{\rm opt} \x) = \frac{\sigma^2}{N} (\x^\T \C^{-1} \x) = \frac{1}{N}\sum_{k=1}^{N} (x_k - x_{k-1})^2  
\end{equation}
(with $x_0 = 0$).  This results agrees with the general Cram\'er-Rao lower
bound which is achieved by the TA MSD with the unit lag time
\cite{Cramer,Voisinne10}.  Moreover, this optimal choice minimizes
simultaneously {\it all} cumulant moments $\kappa_m$ with $m \geq 2$.

In this paper, we extend this analysis to discrete non-centered
Gaussian processes, for instance, in the presence of drift.  We
construct the optimal symmetric matrix $\M_{\rm opt}$ for given mean
vector $\x^0$ and covariance matrix $\C$.  We also derive a simple
explicit formula for the smallest achievable cumulant moment that may
serve as a quality benchmark for other quadratic forms.  In
particular, we compare the optimal matrix $\M_{\rm opt}$ (which
depends on both $\C$ and $\x^0$), to a sub-optimal matrix $\propto
\C^{-1}$ which is independent of $\x^0$ and thus more robust against
uncertainties in $\x^0$ (which is often unknown or difficult to
estimate accurately).  We show that the variance (or higher cumulant
moments) increases by a small amount when the sub-optimal matrix is
used.  Finally, we compare the optimal matrix to the standard
quadratic estimators: TA MSD and TA VACF.  For this purpose, we
consider a discrete fractional Brownian motion (fBm) superimposed with
a constant drift and independent Gaussian noise, as an archetypical
model of anomalous transport affected by measurement artifacts such as
drift and noise.  For this Gaussian process, we compute analytically
the mean and variance of the TA MSD and TA VACF, and compare them to
the optimal matrix.

\section{Distribution of quadratic forms}

In this section, we summarize the basic steps for computing the
distribution of the quadratic form
\begin{equation*}
\chi = \frac12(\x^\T \M \x) ,
\end{equation*}
which is defined by a given symmetric matrix $\M\in \R^N\times \R^N$.
A discrete Gaussian process $\x = \{x_1,...,x_N\} \in\R^N$ is
characterized by its mean vector $\x^0 = \{x_1^0,...,x_N^0\}\in \R^N$
and the covariance matrix $\C$:
\begin{equation*}
x^0_n = \langle x_n\rangle , \qquad \C_{n_1n_2} = \langle x_{n_1} x_{n_2}\rangle - \langle x_{n_1}\rangle \langle x_{n_2}\rangle ,
\end{equation*}
where $\langle \cdots \rangle$ denotes the expectation with respect to
the Gaussian probability density of $\x$:
\begin{equation}
\label{eq:Gaussian}
P_N(\x) = \frac{1}{(2\pi)^{N/2} \sqrt{\det \C}} \exp\biggl[ - \frac12 (\x-\x^0)^\T \C^{-1} (\x-\x^0) \biggr] .
\end{equation}
The characteristic function of the quadratic form $\chi$ is easily
found by regrouping two quadratic forms and computing Gaussian
integrals:
\begin{equation}
\label{eq:phi}
\begin{split}
& \phi(k) \equiv \langle e^{ik\chi}\rangle = \frac{1}{\sqrt{\det (\I - ik \M_\C)}} \times \\
& \exp\biggl[- \frac12 \x^{0,\T} \C^{-1/2} \bigl[\I - (\I - ik\M_\C)^{-1}\bigr]\C^{-1/2} \x^0 \biggr] ,  \\
\end{split}
\end{equation}
where $\M_\C \equiv \C^{1/2}\M\C^{1/2}$, and the inverse and square
root matrices of $\C$ are well defined as the covariance matrix $\C$
is symmetric and positive definite
\footnote{
In the earlier work \cite{Grebenkov11b}, we gave a slightly different
representation in terms of non-symmetric matrix $\M\C$ instead of
$\M_\C = \C^{1/2} \M \C^{1/2}$.  The Sylvester's determinant theorem
\cite{Harville} ensures that $\det(\I - ik\M\C) = \det(\I - ik\M_\C)$
so that both representations are identical for centered Gaussian
processes considered in \cite{Grebenkov11b}. }.
The probability density $p(z)$ of the random variable $\chi$ can be
retrieved through the inverse Fourier transform of $\phi(k)$:
\begin{equation}
\label{eq:pz}
p(z) = \int\limits_{-\infty}^\infty \frac{dk}{2\pi} e^{-ikz} \phi(k) .
\end{equation}
In practice, this computation can be rapidly performed by a fast
Fourier transform.  These basic formulas allow one to study various
quadratic forms of discrete Gaussian processes.  Note that the
probability distribution of quadratic forms of Gaussian processes has
been thoroughly studied in mathematical and physical literature
(see a short overview in \cite{Grebenkov11b}).

The characteristic function $\phi(k)$ can be expressed through the
spectral properties of the matrix $\M_\C$.  Since the square root
matrix $\C^{1/2}$ can be chosen to be symmetric, the matrix $\M_\C$ is
symmetric and thus diagonalizable by an orthogonal matrix ${\bf U}$:
$\M_\C = {\bf U\Lambda U}^\T$, where ${\bf \Lambda}$ is a diagonal
matrix.  One gets therefore
\begin{equation}
\label{eq:phi_k_log}
\ln \phi(k) = -\frac12\sum\limits_{q=1}^{N} \biggl[\ln(1 - ik \lambda_q) + c_q^2 \bigl(1 - (1 - ik\lambda_q)^{-1}\bigr)\biggr],
\end{equation}
where $c_q \equiv (\x^{0,\T} \C^{-1/2} {\bf U})_q$, and $\lambda_q$
are the eigenvalues of $\M_\C$.  The logarithm of $\phi(k)$ is the
generating function for the cumulant moments $\kappa_m \equiv \langle
\chi^m\rangle_c$:
\begin{equation*}
\ln \phi(k) = \sum\limits_{m=1}^\infty \frac{(ik)^m}{m!} \kappa_m .
\end{equation*}
Developing the logarithm in Eq. (\ref{eq:phi_k_log}) into a Taylor
series and identifying the coefficients, one finds
\begin{eqnarray}
\label{eq:kappa}
\kappa_m &=& \frac{m!}{2} \sum\limits_{q=1}^{N} \lambda_q^m \biggl(\frac{1}{m} + c_q^2\biggr) \\ 
\label{eq:kappa2}
&=& \frac{m!}{2} \biggl[\frac{\tr((\M\C)^m)}{m} + \bigl(\x^{0,\T} (\M\C)^{m-1} \M \x^0) \biggr] . 
\end{eqnarray}
For instance, $\kappa_1 = \frac12 \tr(\M\C) + \frac12 (\x^{0,\T} \M
\x^0)$ and $\kappa_2 = \frac12\tr((\M\C)^2) + (\x^{0,\T} \M\C\M\x^0)$
are the mean and the variance of $\chi$, respectively.  The skewness
and kurtosis are also expressed in terms of the cumulant moments as
$\kappa_3/\kappa_2^{3/2}$ and $\kappa_4/\kappa_2^2$, respectively.
Note that the moments $\langle \chi^m\rangle$ can be easily expressed
through the cumulant moments, while the negative-order moments
$\langle \chi^{-\alpha}\rangle$ can be obtained as the Mellin
transform of the characteristic function (see
\cite{Grebenkov11b} for details).  The spectral representations
(\ref{eq:phi_k_log}, \ref{eq:kappa}) extend the analysis of
Ref. \cite{Grebenkov11b} to non-centered Gaussian processes with
nonzero mean $\x_0$ which enters through the coefficients $c_q$.

\section{Optimal quadratic form}

The representation (\ref{eq:kappa}) allows us to tackle the problem of
finding the symmetric matrix $\M_{\rm opt}$ (to be called ``optimal'')
that minimizes a chosen cumulant moment $\kappa_m$ of the random
variable $\chi$, under constraint of the mean value $\kappa_1$ to be
fixed.  For centered Gaussian processes ($\x^0 = 0$), we showed that
the optimal choice is achieved when all eigenvalues $\lambda_q$ of the
matrix $\M_\C$ are identical (and equal to $\lambda = 2\kappa_1/N$),
from which $\M_{\rm opt} = \lambda\C^{-1}$ \cite{Grebenkov11b}.  In
this paper, we extend this result to non-centered Gaussian processes.

A formal solution of the minimization problem for a given cumulant
moment $\kappa_m$ leads to a system of $N\times N$ equations on the
elements of the matrix $\M$:
\begin{equation*}
\frac{\partial}{\partial \M_{jk}} \biggl[\kappa_m + \alpha\biggl(2\kappa_1 - \bigl[\tr(\M\C) + (\x^{0,\T} \M \x^0)\bigr]\biggr)\biggr] = 0 ,
\end{equation*}
where the second term incorporates the constraint (with the Lagrange
multiplier $\alpha$), and $\kappa_m$ is expressed in terms of $\M$
according to Eq. (\ref{eq:kappa2}).  This system is linear only for $m
= 2$.  Although a numerical solution of the system is possible for
small $N$, it does not help to understand the properties of the
optimal solution in general.

The key point of the following analysis is the spectral representation
(\ref{eq:kappa}) of the cumulant moment $\kappa_m$ in terms of
$\lambda_q$ and $c_q$.  The eigenvalues $\lambda_q$ determine the
diagonal matrix ${\bf \Lambda}$, while $c_q$ are the projections of a
given vector $\C^{-1/2}\x^0$ onto the columns of the orthogonal matrix
$\U$.  The constrained minimization of $\kappa_m$ is equivalent to
unconstrained minimization of the function
\begin{equation}
\label{eq:f}
\begin{split}
f & = \sum\limits_q \lambda_q^m (1 + m c_q^2) \\
& + \alpha \biggl(2\kappa_1 - \sum\limits_q \lambda_q (1 + c_q^2)\biggr) + \beta \biggl(\sum\limits_q c_q^2 - \gamma \biggr).\\
\end{split}
\end{equation}
with respect to $\lambda_q$ and $c_q$.  Here $\alpha$ and $\beta$ are
two Lagrange multipliers that implement two constraints:
\begin{eqnarray}
\label{eq:constraint_lambda}
\sum\limits_q \lambda_q (1 + c_q^2) &=& 2\kappa_1 , \\
\label{eq:constraint_c}
\sum\limits_q c_q^2 &=&  
(\x^{0,\T} \C^{-1} \x^0) \equiv \gamma . 
\end{eqnarray}
The first constraint eliminates a trivial solution $\M = 0$ that would
minimize all the cumulant moments.  The second relation accounts for
the orthogonality of the matrix $\U$.

In what follows, we consider $m \geq 2$ to be even, in order to ensure
that the function $f$ is bounded from below and thus admits a minimum.
Setting the derivatives of $f$ with respect to $\lambda_q$ and $c_q$
to zero yields two sets of equations:
\begin{eqnarray}
m \lambda_q^{m-1} (1 + mc_q^2) - \alpha(1 + c_q^2) &=& 0, \\
(m \lambda_q^m - \alpha \lambda_q + \beta)c_q &=& 0 ,
\end{eqnarray}
which are completed by the constraints (\ref{eq:constraint_lambda},
\ref{eq:constraint_c}).  The first equation yields
\begin{equation}
\label{eq:lambda_q}
\lambda_q = \lambda \left(\frac{1+c_q^2}{1+mc_q^2}\right)^{\frac{1}{m-1}} , \qquad
\lambda \equiv \left(\frac{\alpha}{m}\right)^{\frac{1}{m-1}} .
\end{equation}
The second equation admits two options: 

(i) $c_q^2 = 0$, for which $\lambda_q = \lambda$ according to
Eq. (\ref{eq:lambda_q});

(ii) $c_q^2 > 0$, in which case one has to solve the equation
\begin{equation}
\label{eq:auxil1}
m \lambda_q^m - \alpha \lambda_q + \beta = 0. 
\end{equation}
Since $\alpha$ and $\beta$ are constants (to be determined), solutions
of this equation have the same form for all $q$.  In general, the
$m$-th order polynomial in Eq. (\ref{eq:auxil1}) has $m$
(complex-valued) roots.  In Appendix \ref{sec:oneroot}, we show that
only one root of Eq. (\ref{eq:auxil1}) is compatible with
Eq. (\ref{eq:lambda_q}).

There is still a freedom to choose one of two above options for every
$q$.  Let $Q$ denote the number of nonzero coefficients $c_q^2$.  The
statement of Appendix \ref{sec:oneroot} implies that $\lambda_1 =
... = \lambda_Q$ and thus $c_1^2 = ... = c_Q^2 > 0$.  For the
remaining indices $q = Q+1,...,N$, one has $c_q^2 = 0$ and $\lambda_q
= \lambda$.  Substituting these relations in
Eqs. (\ref{eq:constraint_lambda}, \ref{eq:constraint_c}) yields $c_1^2
= \gamma/Q$ and
\begin{equation*}
2\kappa_1 = \sum\limits_q \lambda_q (1+c_q^2) = Q\lambda_1 (1 + \gamma/Q) + (N-Q) \lambda ,
\end{equation*}
from which
\begin{equation*}
\lambda_1 = \frac{2\kappa_1 - (N-Q)\lambda}{Q+\gamma} .
\end{equation*}
Comparing this relation with Eq. (\ref{eq:lambda_q}) leads to
\begin{equation}
\label{eq:lambda_const}
\lambda = \frac{2\kappa_1}{(Q+\gamma)\left(\frac{Q+\gamma}{Q+m\gamma}\right)^{\frac{1}{m-1}} + (N-Q)} ,
\end{equation}
so that
\begin{equation}
\label{eq:lambda}
\lambda_q = \lambda \times \begin{cases} \left(\frac{Q+\gamma}{Q+m\gamma}\right)^{\frac{1}{m-1}} , \quad q = 1,...,Q, \cr
~ \hskip 6mm  1, \hskip 15mm q = Q+1,...,N. \end{cases}
\end{equation}
In other words, the fact that only one solution of
Eq. (\ref{eq:auxil1}) is admissible allows us to omit resolution of
this equation.

For centered processes (i.e., $\x^0 = 0$ and $\gamma = 0$), one
retrieves $\lambda_q = \lambda = 2\kappa_1/N$ that minimized all the
cumulant moments.  In contrast, when $\gamma > 0$, the optimal
solution $\lambda_q$ depends on the order $m$ of the moment to be
minimized.  

The $n$-th cumulant moment for the optimal form reads according to
Eq. (\ref{eq:kappa}) as
\begin{equation}
\kappa_n^{\rm opt} = \frac{(n-1)!}{2} \lambda^n \biggl[\left(\frac{Q+\gamma}{Q+m\gamma}\right)^{\frac{n}{m-1}} (Q + n\gamma) + (N-Q)\biggr] ,
\end{equation}
where $n$ may be different from $m$.  When $n = m$, one gets
\begin{equation}
\label{eq:kappa_opt}
\kappa_m^{\rm opt} = \frac{(m-1)!}{2} \lambda^{m-1} 2\kappa_1 .
\end{equation}
In particular, one has for $m = n = 2$
\begin{equation}
\label{eq:kappa2_opt}
\kappa_2^{\rm opt} = \frac{2\kappa_1^2}{N} \biggl(1 + \frac{\gamma^2}{N(Q+2\gamma)}\biggr)^{-1} .
\end{equation}

The function $\kappa_m^{\rm opt}$ monotonously increases with $Q$
because
\begin{equation*}
\frac{\partial \lambda}{\partial Q} = \frac{\lambda^2}{2\kappa_1} \biggl[1 - \left(\frac{Q+\gamma}{Q+m\gamma}\right)^{\frac{1}{m-1}} 
\frac{Q + \gamma(m+1)}{Q+m\gamma}\biggr] \geq 0  ,
\end{equation*}
that follows from the inequality
\begin{equation*}
(1 + z(m+1))^{m-1} (1+z) \leq (1+mz)^m ,
\end{equation*}
where $z = \gamma/(Q+\gamma)$ lies between $0$ and $1$.  As a
consequence, the minimum of the $m$-th cumulant moment is reached for
$Q = 1$.  In particular, Eqs. (\ref{eq:lambda_const},
\ref{eq:kappa_opt}) with $Q = 1$ give a simple explicit formula for
the smallest achievable cumulant moment.

Note also that nonzero mean vector (e.g., a drift) always diminishes
the optimal cumulant moment $\kappa_m^{\rm opt}$ because
\begin{equation*}
\frac{\partial \lambda}{\partial \gamma} = - \frac{m\gamma \lambda^2}{2\kappa_1} ~ \frac{(Q+\gamma)^{\frac{1}{m-1}}}{(Q+m\gamma)^{\frac{m}{m-1}}} < 0 .
\end{equation*}
In addition, the parameter $\lambda$ and thus the optimal moment
$\kappa_m^{\rm opt}$ go to $0$ as $\gamma\to \infty$ according to
Eqs. (\ref{eq:lambda_const}, \ref{eq:kappa_opt}).  In fact, the
contribution of the mean vector $\x^0$ strongly dominates over random
fluctuations in this limit.  We emphasize again that this statement
remains correct only under the constraint of fixed mean value.

When the mean vector $\x^0$ and covariance matrix $\C$ are known, the
optimal matrix $\M_{\rm opt}$ can be constructed as follows.  First,
one sets the vector $\u_1 = \C^{-1/2} \x^0/\sqrt{\gamma}$ and then
chooses $N-1$ orthonormal vectors $\u_q$ ($q=2,...,N$) that are all
orthogonal to $\u_1$ so that $c_q = (\u_q \cdot \C^{-1/2} \x^0) =
(\u_q \cdot \sqrt{\gamma}\u_1) = 0$.  By construction, $c_1 = (\u_1
\cdot \C^{-1/2} \x^0) = (\x^{0,\T} \C^{-1} \x^0)/\sqrt{\gamma} =
\sqrt{\gamma}$.  The vectors $\u_q$ form the orthogonal matrix $\U$.
After that, one constructs a diagonal matrix ${\bf
\Lambda}$ which has the first element $\lambda_1 = \lambda
\bigl(\frac{1+\gamma}{1+m\gamma}\bigr)^{\frac{1}{m-1}}$ and the other
diagonal elements $\lambda_q = \lambda$, with $\lambda$ from
Eq. (\ref{eq:lambda_const}) with $Q = 1$.  One gets therefore
\begin{equation*}
[\U {\bf \Lambda} \U^+]_{j,k} = \lambda \delta_{j,k} + \frac{\lambda_1 - \lambda}{\gamma} (\C^{-1/2} \x^0)_j (\C^{-1/2} \x^0)_k ,
\end{equation*}
from which the identity $\M = \C^{-1/2} \U {\bf \Lambda} \U^\T
\C^{-1/2}$ yields the optimal matrix $\M_{\rm opt}$ as
\begin{equation}
\label{eq:Mopt}
[\M_{\rm opt}]_{j,k} = \lambda [\C^{-1}]_{j,k} - \frac{\eta}{\gamma} (\C^{-1} \x^0)_j (\C^{-1} \x^0)_k ,
\end{equation}
with 
\begin{equation}
\eta \equiv \lambda - \lambda_1 = \lambda \biggl(1 - \biggl(\frac{1+\gamma}{1+m\gamma}\biggr)^{\frac{1}{m-1}}\biggr) .
\end{equation}
One can see how the mean vector $\x^0$ modifies the optimal matrix
through the prefactor $\lambda$ [given by Eq. (\ref{eq:lambda_const})]
and the second term in Eq. (\ref{eq:Mopt}).

\subsection*{Sub-optimal matrix}

A limitation of the above approach is the need for knowing the
covariance matrix $\C$ and the mean vector $\x^0$.  In particular, the
resulting optimal matrix $\M_{\rm opt}$ depends not only on the
structure of $\C$ and $\x^0$, but also on the parameters (e.g., the
drift coefficient) which are often unknown and have to be inferred.
For inference purposes, one needs to find such a matrix $\M_{\rm sub}$
which may be sub-optimal but more robust against changes of the
parameters.  Quite remarkably, one can check that the optimal moment
$\kappa_m^{\rm opt}$ from Eq. (\ref{eq:kappa_opt}) weakly depends on
$Q$.  In the ``worst'' case $Q = N$, the diagonal matrix ${\bf
\Lambda}$ is simply proportional to the identity matrix, ${\bf
\Lambda} = \frac{2\kappa_1}{N+\gamma} \I$, so that the related
sub-optimal matrix becomes
\begin{equation}
\M_{\rm sub} = \frac{2\kappa_1}{N+\gamma} \C^{-1} .
\end{equation}
The crucial simplification here is that the mean vector $\x^0$ enters
only through the prefactor $1/(N+\gamma)$ in front of $\C^{-1}$.  In
other words, the structure of the matrix $\M_{\rm sub}$ does not
depend on the particular structure of the mean vector.  Although the
matrix $\M_{\rm sub}$ is less optimal than $\M_{\rm opt}$, the
difference between the $m$-th cumulant moments for both matrices is
small.  For instance, when $m = 2$, the difference between the
variances of the optimal and sub-optimal quadratic forms can be found
from Eq. (\ref{eq:kappa2_opt}) as
\begin{equation*}
\kappa_2^{\rm sub} - \kappa_2^{\rm opt} = \frac{2\kappa_1^2 (N-1) \gamma^2}{(N^2+2\gamma N+\gamma^2)(N + 2\gamma N + \gamma^2)} .
\end{equation*}
This difference vanishes at $\gamma = 0$ and $\gamma\to\infty$,
attending the maximum value $\approx 0.09(2\kappa_1^2/N)$ at $\gamma
\approx \frac{\sqrt{5}-1}{2}~ N$ (for large $N$).  One concludes that
the use of the sub-optimal matrix $\M_{\rm sub}$ instead of $\M_{\rm
opt}$ may increase the variance by at most $\approx 10\%$.

\section{Fractional Brownian motion}
\label{sec:fBm}

As we discussed earlier, the TA MSD with the lag time $n=1$ is the
optimal quadratic functional for Brownian motion.  The simplicity of
the TA MSD made this quadratic functional broadly employed for the
analysis of more sophisticated processes such as anomalous diffusions.
Various types of motion in living cells, biological tissues and
mineral samples were observed and analyzed: restricted, obstructed and
hindered diffusion, directed motion, anomalous diffusion, diffusion
through traps,
etc. \cite{Saxton97b,Saxton93,Metzler09,Bouchaud90,Metzler00,Grebenkov07}.
One may wonder how strongly the efficiency of the TA MSD for other
Gaussian processes is reduced as compared to the optimal quadratic
form.  In other words, it is instructive to compare the variance of
the TA MSD for anomalous diffusion to the optimal variance
$\kappa_2^{\rm opt}$ from Eq. (\ref{eq:kappa2_opt}).  Given that the
optimal variance was obtained under the constraint of fixed mean
$\kappa_1$, it is convenient to consider the ratio
$\kappa_2/\kappa_1^2$.  Note that this ratio for TA MSD was also
called ``ergodicity breaking parameter'' and thoroughly studied for
anomalous diffusions
\cite{Deng09,Jeon10,Jeon11,Burov11}.  In particular, the behavior of
this ratio in the limit $N\to\infty$ would tell about ergodic
properties of the system: if this ratio vanishes as $N\to\infty$, the
time average over infinitely long trajectory is equivalent to the
ensemble average (the ergodic property); in turn, the nonzero limit
indicates weak ergodicity breaking which was observed and investigated
for different kinds of continuous-time random walks (see
\cite{Burov11} and references therein).

As an archetypical model of anomalous diffusion, we consider a
(discrete) fractional Brownian motion with the Hurst exponent $0 < H <
1$ (its continuous version was introduced by Kolmogorov
\cite{Kolmogorov40} and later by Mandelbrot and van Ness
\cite{Mandelbrot68}).  We also add a constant drift $\mu$ and an
independent Gaussian noise with mean zero and variance $\ve^2$ that
may account for some measurement artifacts
\cite{Berglund10,Michalet10,Michalet12}.  The resulting process
(starting from $0$) is still Gaussian and thus fully characterized by
the mean vector $\x^0 = \mu (1,2,...,N)^\T$ and the covariance matrix
\begin{equation}
\label{eq:C_fBm}
\C_{n_1,n_2} = \frac{\sigma^2}{2} \bigl(n_1^{2H} + n_2^{2H} - |n_1-n_2|^{2H}\bigr) + \ve^2 \delta_{n_1,n_2}.
\end{equation}
The fBm is persistent (with positive correlations between steps) or
anti-persistent (with negative correlations between steps) for $H >
1/2$ and $H < 1/2$, respectively.  Finally, one retrieves Brownian
motion at $H = 1/2$.

\subsection{Optimal variance}

In contrast to Brownian motion, the inverse matrix $\C^{-1}$ is not
known explicitly for fBm.  Setting $\ve = 0$ for simplicity, we check
numerically that the coefficient $\gamma$ defined by
Eq. (\ref{eq:constraint_c}) behaves as $\gamma \approx
\frac{\mu^2}{\sigma^2} N^{2(1-H)}(c_H + O(1/N))$, where the constant
$c_H$ is close to $1$ and weakly dependent on $H$ when $H$ is not too
small.  According to Eq. (\ref{eq:kappa2_opt}), we get the optimal
ratio
\begin{equation}
\frac{\kappa_2^{\rm opt}}{2\kappa_1^2} \simeq \frac{1}{N} \left(1 + \frac{\mu^4 c_H^2}{\sigma^4} 
\frac{N^{3-4H}}{1 + 2\frac{\mu^2}{\sigma^2} N^{2(1-H)}}\right)^{-1} .
\end{equation}
Two limiting cases of small and large drift $\mu$, as compared to
$\sigma N^{H-1}$, yield
\begin{equation}
\frac{\kappa_2^{\rm opt}}{2\kappa_1^2} \simeq \begin{cases}
\frac{1}{N} \bigl(1 + \frac{\mu^4 c_H^2}{\sigma^4} N^{3-4H}\bigr)^{-1} \quad (\mu \ll \sigma N^{H-1}) , \cr
\frac{1}{N} \bigl(1 + \frac{\mu^2 c_H^2}{2\sigma^2} N^{1-2H}\bigr)^{-1} \quad (\mu \gg \sigma N^{H-1}) . \end{cases}
\end{equation}
Since $H < 1$, the ``border'' $\sigma N^{H-1}$ between two asymptotic
limits decreases with $N$.  In addition, the drift ``correction'' may
be either enhanced, or damped with $N$ for different values of $H$.
For instance, when $H > 1/2$, the second term in the second relation
decreases, i.e., the role of the drift progressively diminishes.  In
turn, when $H < 1/2$, the drift changes significantly the properties
of the optimal quadratic form.

\subsection{Comparison with TA MSD}

The TA MSD with the lag time $n$ over a sample of length $N$ is
defined as a moving average
\begin{equation}
\label{eq:TAMSD}
\chi = \frac{1}{N-n} \sum\limits_{k=1}^{N-n} (x_{k+n} - x_k)^2 
\end{equation}
(note that this analysis is also applicable to multi-dimensional
processes for which the TA MSD is simply the sum of TA MSDs for each
component).  One can notice that Eq. (\ref{eq:TAMSD}) with the lag
time $n = 1$ is slightly different from the optimal one defined by
Eq. (\ref{eq:TAMSD0}).  In fact, the latter contains an additional
term $(x_1 - x_0)^2 = x_1^2$ and the prefactor $1/N$ instead of
$1/(N-1)$.  Nevertheless, we keep the definition (\ref{eq:TAMSD})
which is commonly used.  All the presented results can be recomputed
for Eq. (\ref{eq:TAMSD0}) as well.

After lengthy computation, we obtain the mean and variance of the TA
MSD for the discrete fBm with a constant drift and independent
Gaussian noise:
\begin{equation}
\label{eq:kappa1_TAMSD}
\kappa_1 = \sigma^2 n^{2H} + \mu^2 n^2 + 2\ve^2 ,
\end{equation}
and
\begin{equation}
\label{eq:kappa2_TAMSD}
\begin{split}
\kappa_2 & = \frac{2}{N-n} \biggl[ \sigma^4 c_{\sigma,\sigma} n^{4H} + 2\mu^2\sigma^2 c_{\sigma,\mu} n^{2+2H}  \\
& + 4\mu^2 \ve^2 c_{\mu,\ve} n^2 + 4\ve^2\sigma^2 c_{\sigma,\ve} n^{2H} + 4\ve^4 c_{\ve,\ve}  \biggr],  \\
\end{split}
\end{equation}
where
\begin{equation*}
\begin{split}
c_{\sigma,\sigma}(n,N) & = 1 + S_H^{(2)}(n,N),  \\
c_{\sigma,\mu}(n,N) & = 1 + S_H^{(1)}(n,N),  \\
c_{\mu,\ve}(n,N) & = \begin{cases} \frac{n}{N-n} \quad (n < N/2) , \cr 1 \hskip 9mm (n \geq N/2) , \end{cases} \\
c_{\ve,\ve}(n,N) & = \begin{cases} \frac32 - \frac{n}{2(N-n)} \quad (n < N/2), \cr 1  \hskip 19mm (n \geq N/2), \end{cases} \\
c_{\sigma,\ve}(n,N) & = \begin{cases} 1 - (2^{2H-1}-1)\frac{N-2n}{N-n} \quad (n < N/2) ,\cr 1  \hskip 35mm  (n\geq N/2), \end{cases} \\
\end{split}
\end{equation*}
and
\begin{equation}
\label{eq:SH}
\begin{split}
S_H^{(m)}(n,N) & \equiv \frac{1}{m~ n^{2Hm} (N-n)} \hspace*{-1mm}\sum\limits_{k=1}^{N-n-1} \hspace*{-1mm} (N-n-k) \\
& \times \biggl(|k-n|^{2H} + |k+n|^{2H} - 2k^{2H}\biggr)^m . \\
\end{split}
\end{equation}

The general formulas (\ref{eq:kappa1_TAMSD}, \ref{eq:kappa2_TAMSD})
allow one to study the dependence of the mean and variance of the TA
MSD on different parameters of the studied process, namely, $H$,
$\sigma$, $\mu$ and $\ve$.  In particular, we will investigate the
behavior of the ratio $\kappa_2/(2\kappa_1^2)$ for different
situations.

\subsubsection*{Brownian motion}

For $H = 1/2$, the sum (\ref{eq:SH}) can be computed explicitly, from
which
\begin{equation*}
c_{\sigma,\mu}(n,N) = \begin{cases} \frac{1 + 3Nn - 4n^2}{3(N-n)} \hskip 17mm (n<N/2) ,\cr
\frac{1 + 3N(N-n) - 4(N-n)^2}{3n}  \quad (n\geq N/2) , \end{cases}
\end{equation*}
and $c_{\sigma,\sigma}(n,N) = (N-n) F_{n,N}$, where the prefactor
$F_{n,N}$ was first derived by Qian {\it et al.} \cite{Qian91}:
\begin{equation*} 
F_{n,N} = \begin{cases} \frac{4n^2 N - 5n^3 + 2N - n}{6n(N-n)^2}  \hskip 29mm (n \leq N/2), \cr
\frac{(N-n)^3 - 4n(N-n)^2 + 6n^2(N-n) + 5n - N}{6n^2(N-n)} \hskip 2mm (n \geq N/2). \end{cases} 
\end{equation*}
The prefactor $F_{n,N}$ is an increasing function of $n$ which ranges
from $\frac{1}{N-1}$ at $n = 1$ to $1$ at $n = N-1$.

When there is no drift ($\mu = 0$), Eq. (\ref{eq:kappa2_TAMSD}) is
reduced to
\footnote{
Note that the prefactor $2$ was erroneously omitted in Eq. (26) of
Ref. \cite{Grebenkov11b}.  In fact, the original expression for
$F_{n,N}$ was derived by Qian {\it et al.} \cite{Qian91} for
two-dimensional Brownian motion.  In the one-dimensional case, the
variance is twice larger. }
%
\begin{equation*}
\kappa_2 = 2n^2 F_{n,N} + \frac{8n \ve^2}{N-n} + \frac{8\ve^4}{N-n} \times \begin{cases}  \frac{3N-4n}{2(N-n)} ~~ (n \leq N/2) , \cr
 1    \hskip 11.5mm  (n > N/2) . \end{cases}
\end{equation*}
One can check that the ratio $\kappa_2/(2\kappa_1^2)$ is an increasing
function of $n$.  As expected, the ratio is minimal for $n = 1$, for
which
\begin{equation*}
\frac{\kappa_2}{2\kappa_1^2} = \frac{1}{N-1} \biggl(1 + 2\ve^4 \frac{1 - \frac{3}{N-1}}{(1+2\ve^2)^2}\biggr) .
\end{equation*}
The noise monotonously increases the ratio, from the (almost) optimal
value $\frac{1}{N-1}$ to approximately $\frac32~\frac{1}{N-1}$ in the
limit of very large noises.  In turn, the optimal quadratic form
obtained by inverting the matrix $\C$ from Eq. (\ref{eq:C_fBm}) would
give the minimal ratio $1/N$.

As we mentioned earlier, the inverse of the covariance matrix $\C$
from Eq. (\ref{eq:C_fBm}) with $H = 1/2$ and $\ve = 0$ (no noise) can
be found explicitly, and $\C^{-1}$ determines the TA MSD with the unit
lag time according to Eq. (\ref{eq:TAMSD0}).  As a consequence, one
has $(\C^{-1} \x_0)_j = \mu \delta_{j,N}/\sigma^2$ so that $\gamma =
\mu^2 N/\sigma^2$, and the optimal matrix from Eq. (\ref{eq:Mopt})
gets an explicit form
\begin{equation}
[\M_{\rm opt}]_{j,k} = \lambda [\C^{-1}]_{j,k} - \frac{\eta}{N\sigma^2} ~\delta_{j,N} ~\delta_{k,N} .
\end{equation}
One can see that the drift term modifies the prefactor $\lambda$ in
front of $\C^{-1}$ and also changes the last element of the matrix
$\M_{\rm opt}$.   We conclude 
\begin{equation*}
\frac12 (\x^\T \M_{\rm opt} \x) =
\frac{\lambda}{2\sigma^2}\sum_{n=1}^{N} (x_n - x_{n-1})^2 - \frac{\eta}{2N\sigma^2} x_n^2 .
\end{equation*}
This explicit result is specific to discrete Brownian motion.

\begin{figure}
\begin{center}
\includegraphics[width=80mm]{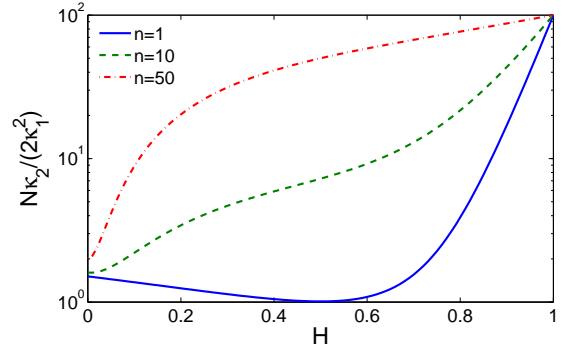}
\end{center}
\caption{
(Color online) The ratio $\kappa_2/(2\kappa_1^2)$ for the TA MSD of
the discrete fBm versus the Hurst exponent $H$ (with $N = 100$,
$\sigma = 1$, $\mu = \ve = 0$), with different lag times: $n = 1$
(blue solid line), $n = 10$ (green dashed line) and $n = 50$ (red
dash-dotted line).  This ratio is normalized by optimal value $1/N$. }
\label{fig:k2k1_H}
\end{figure}

\subsubsection*{Fractional Brownian motion}

When $\mu = \ve = 0$, one gets
\begin{equation}
\label{eq:k2k1_fBm}
\frac{\kappa_2}{2\kappa_1^2} = \frac{c_{\sigma,\sigma}}{N-n} .
\end{equation}
This ratio monotonously increases with $n$, i.e., for any $H$, the
minimal ratio is achieved for $n = 1$, as expected.  Setting $n=1$,
one checks that the positive sum $S_H^{(2)}(1,N)$ vanishes at $H =
1/2$ that minimizes the ratio in Eq. (\ref{eq:k2k1_fBm}).  This is
expected as the TA MSD with the unit lag time is optimal for Brownian
motion.  For $H < 1/2$, the largest ratio corresponds to the limit $H
= 0$ and is equal to $\frac{1}{N-1}
\bigl(\frac{3}{2} - \frac{1}{2(N-1)}\bigr)$ which is $50\%$ larger
than the value $\frac{1}{N-1}$ at $H = 1/2$.  In contrast, this ratio
rapidly increases for $H > 1/2$, attending the value $1$ at $H = 1$.
The behavior of the ratio $\kappa_2/(2\kappa_1^2)$ is illustrated on
Fig. \ref{fig:k2k1_H} for three lag times.  We conclude that the TA
MSD can still be applied to the analysis of subdiffusive fBm with $H <
1/2$, while other quadratic functionals would significantly outperform
the TA MSD for superdiffusive fBm with $H > 1/2$.

It is also instructive to analyze the dependence of
$\kappa_2/(2\kappa_1^2)$ on the sample length $N$.  For $n = 1$, the
sum $S_H^{(m)}(1,N)$ asymptotically behaves as $\propto N^{2m(H-1)+1}$
for large $N$.  As a consequence, two different situations have to be
distinguished: for $H < 3/4$, the coefficient $c_{\sigma,\sigma}(1,N)$
converges to a constant as $N\to\infty$, while for $H > 3/4$, it
diverges as $\propto N^{4(H-3/4)}$ (for $H = 3/4$, the divergence is
logarithmic).  In other words, for $H < 3/4$, the statistical
uncertainty remains of the order of $N^{-1}$ independently of $H$; in
turn, for $H > 3/4$, the decrease rate is dependent on $H$ and slower:
$\propto N^{-4(1-H)}$.  This behavior for continuous-time fBm was
derived analytically by Deng and Barkai \cite{Deng09} (see also
\cite{Jeon10}).

\begin{figure}
\begin{center}
\includegraphics[width=80mm]{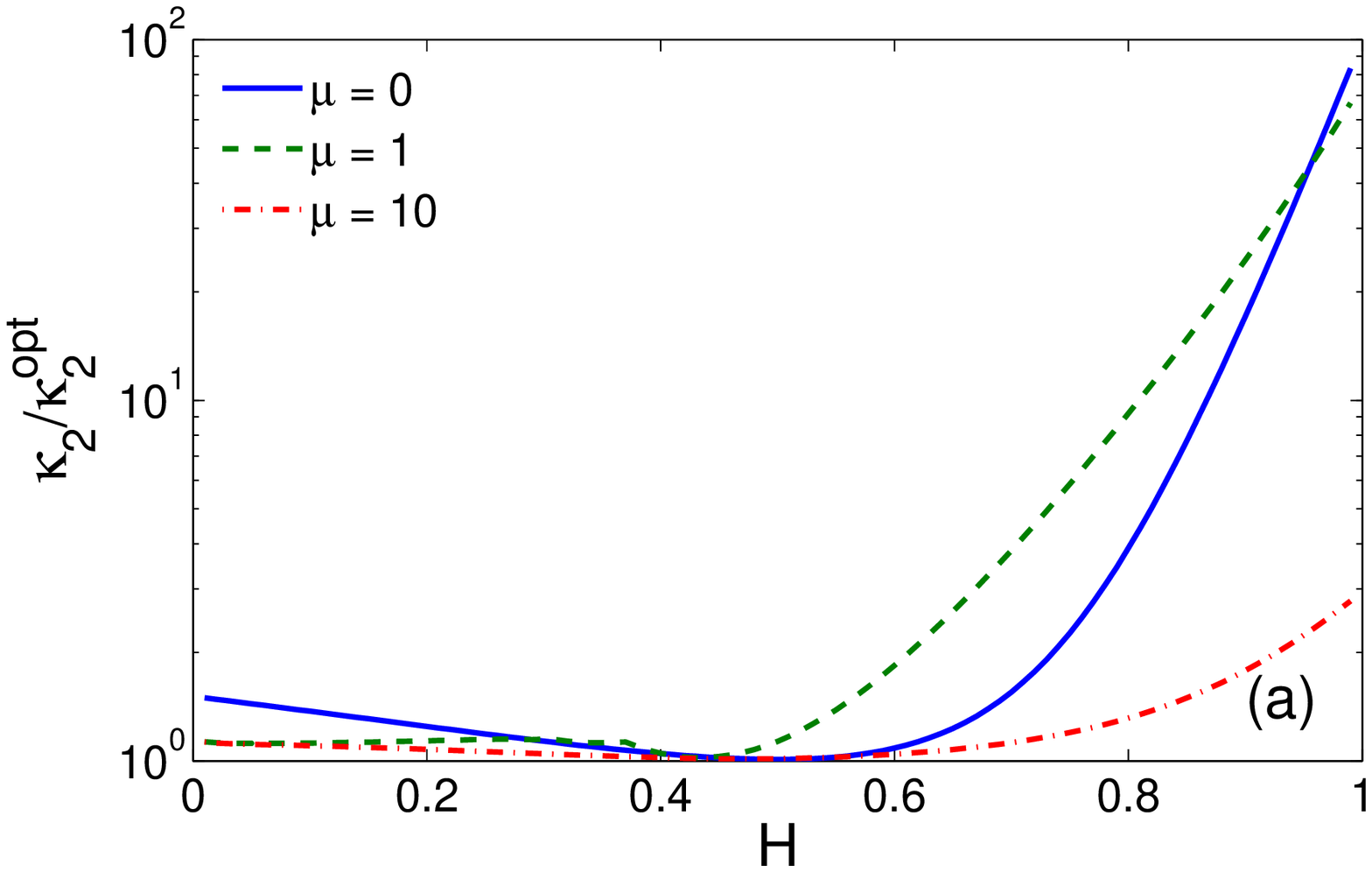}
\includegraphics[width=80mm]{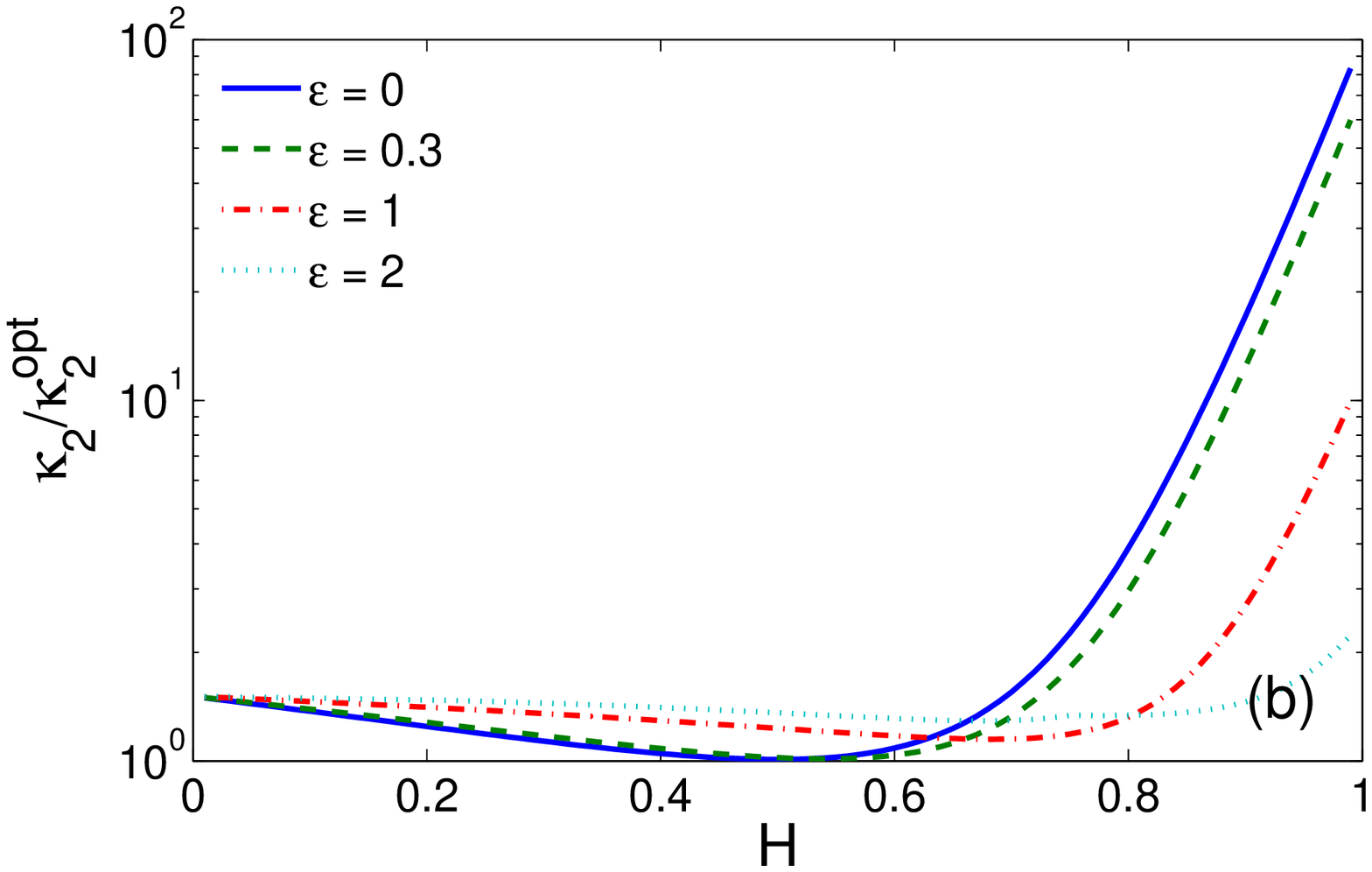}
\end{center}
\caption{
(Color online) The ratio $\kappa_2/\kappa_2^{\rm opt}$ for the TA MSD
of the discrete fBm versus the Hurst exponent $H$ (with $N = 100$,
$\sigma = 1$): {\bf (a)} for three drifts $\mu$ (without noise, $\ve =
0$); {\bf (b)} for four noise levels $\ve$ (without drift, $\mu=0$).
Here, one uses the minimal value of $\kappa_2$ over all lag times.}
\label{fig:MSD_ratio}
\end{figure}

The effect of drift and noise onto the variance $\kappa_2$ of the TA
MSD may be quite sophisticated.  First, the ratio
$\kappa_2/\kappa_1^2$ is not necessarily minimal at the lag time $n =
1$.  For this reason, we consider the minimal value of the ratio
$\kappa_2/(2\kappa_1^2)$ over all lag times $n$ from $1$ to $N-1$.
This ratio is then normalized by the optimal ratio $\kappa_2^{\rm
opt}/(2\kappa_1^2)$ that is equivalent to plotting
$\kappa_2/\kappa_2^{\rm opt}$.  By construction, the latter ratio is
always greater than $1$.  Small deviations from $1$ would mean nearly
optimal inference power of the TA MSD.  Figure \ref{fig:MSD_ratio}
illustrates the optimality of the TA MSD for fBm altered by drift and
noise.  When drift coefficient $\mu$ or noise level $\ve$ are large,
the ratio $\kappa_2/\kappa_2^{\rm opt}$ is getting smaller and weakly
dependent on $H$.  This is not surprising as drift or noise dominates
over fBm at large $\mu$ or $\ve$.  Interestingly, the drift shifts the
minimum of the ratio towards $H < 1/2$ (Fig. \ref{fig:MSD_ratio}a),
while the noise shifts the minimum towards $H > 1/2$
(Fig. \ref{fig:MSD_ratio}b).

\subsection{Comparison with TA VACF}

The velocity auto-correlation function, $\langle \dot{X}(t_1)
\dot{X}(t_2)\rangle$, provides a direct measure of correlations
between elementary displacements of the process.  For the analysis of
individual trajectories, the ensemble average has to be replaced by
the time average of $\dot{X}(t_0+t) \dot{X}(t_0)$ with a fixed lag
time $t$ and $t_0$ sliding along the sample trajectory.  In practice,
the time derivative (denoted by dot) is approximated by a finite
difference between the neighboring positions so that the discrete TA
VACF is
\begin{equation}
\chi = \frac{1}{N-n-1} \sum\limits_{k=1}^{N-n-1} (x_{k+n+1} - x_{k+n}) (x_{k+1}-x_k) 
\end{equation}
(the factor $1/\delta^2$ is omitted for the sake of simplicity).
Similarly to TA MSD, this expression defines a quadratic form $\frac12
(\x^{\T} \M \x)$ associated with a symmetric matrix $\M$ whose
elements can be written explicitly.

We compute the mean and variance of the TA VACF for the discrete
fractional Brownian motion with drift and noise:
\begin{equation}
\kappa_1 = \frac{\sigma^2}{2} \biggl[(n+1)^{2H} + (n-1)^{2H} - 2n^{2H}\biggr] + \mu^2 - \ve^2 \delta_{n,1}, 
\end{equation}
and
\begin{equation}
\kappa_2 = \frac{\sigma^4 \tilde{c}_{\sigma,\sigma} + \sigma^2 \ve^2 \tilde{c}_{\sigma,\ve} + \ve^4 \tilde{c}_{\ve,\ve}
+ \mu^2 \sigma^2 \tilde{c}_{\sigma,\mu} + \mu^2 \ve^2 \tilde{c}_{\mu,\ve}}{N-n-1}  ,
\end{equation}
with the explicit but lengthy formulas for the coefficients (dependent
on $n$ and $N$) provided in Appendix \ref{sec:VACF}.

\subsubsection*{Brownian motion}

For Brownian motion ($H = 1/2$), the above expressions are simplified:
$\tilde{c}_{\sigma,\sigma} = 1$, $\tilde{c}_{\sigma,\ve} = 4$, and 
\begin{equation*}
\tilde{c}_{\sigma,\mu} = 2 + \begin{cases} 2\frac{N-2n-1}{N-n-1} \quad (n< N/2), \cr 0 , \hskip 14mm (n\geq N/2) .\end{cases}
\end{equation*}

When there is no drift and noise, the mean TA VACF is zero because all
the displacements of Brownian motion are independent.  Since the
variance $\kappa_2 = \frac{1}{N-n-1}$ is finite, the ratio
$\kappa_2/(2\kappa_1^2)$ is infinite.  Clearly, the TA VACF (as well
as any other quadratic form with zero mean) is far from being optimal
because a measured empirical value does not allow one to infer any
parameter of the process.

In the presence of drift, the mean value $\kappa_1 = \mu^2$ (for $n >
1$) allows one to estimate the drift coefficient.  Neglecting noise,
one gets
\begin{equation*}
\frac{\kappa_2}{2\kappa_1^2} = \frac{\sigma^2}{N-n-1} \biggl(1 + \frac{\sigma^2}{2\mu^2} + 
\max\biggl\{0 ,1 - \frac{n}{N-n-1}\biggr\} \biggr) ,
\end{equation*}
which is minimal at $n = 1$: $\frac{\kappa_2}{2\kappa_1^2}\approx
\frac{\sigma^2}{N-2}\bigl(2+\frac{\sigma^2}{2\mu^2}\bigr)$.

\subsubsection*{Fractional Brownian motion}

\begin{figure}
\begin{center}
\includegraphics[width=80mm]{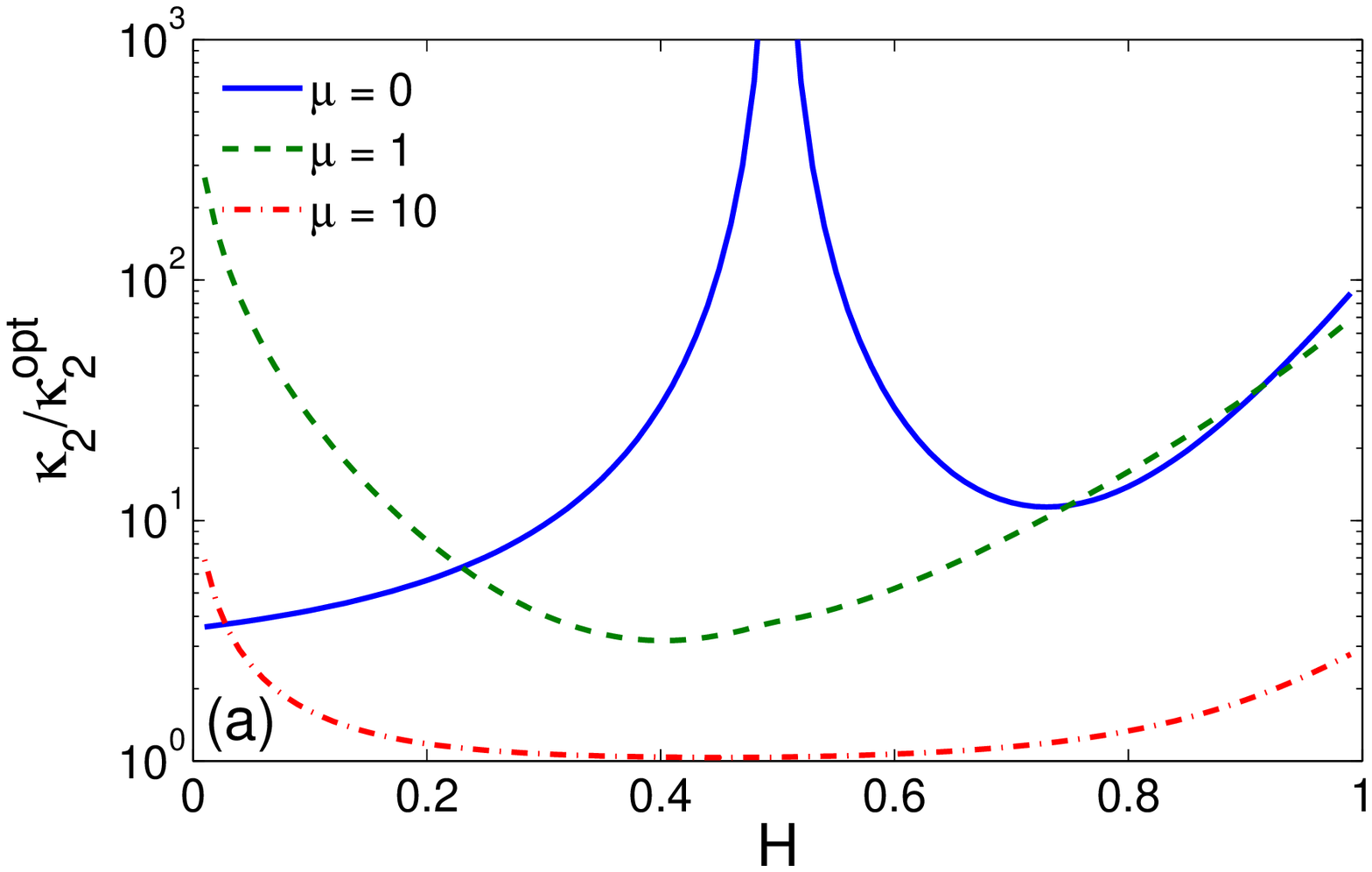}
\includegraphics[width=80mm]{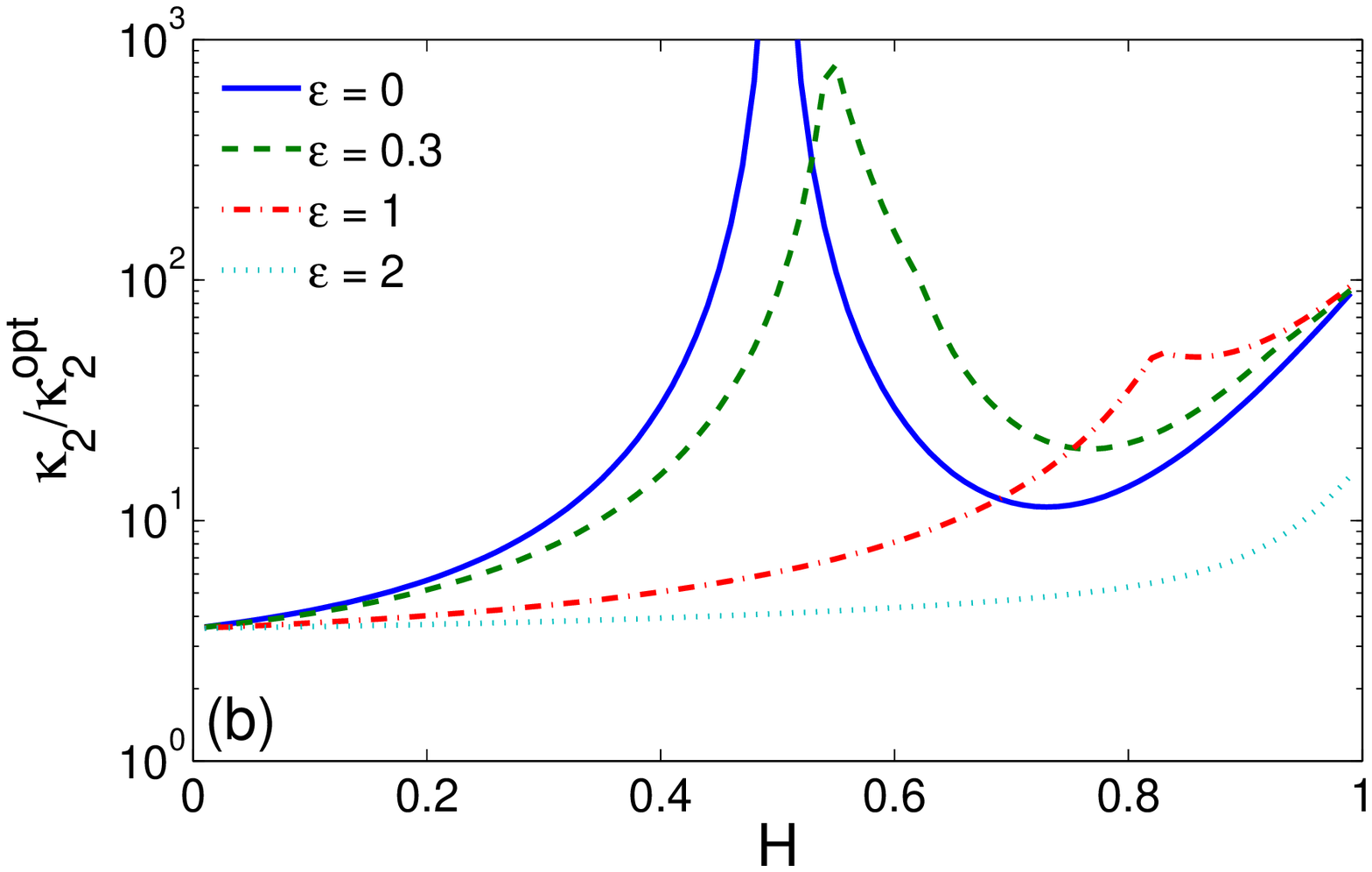}
\end{center}
\caption{
(Color online) The ratio $\kappa_2/\kappa_2^{\rm opt}$ for the TA VACF
of the discrete fBm of the Hurst exponent $H$ (with $N = 100$, $\sigma
= 1$).  {\bf (a)} For three drifts $\mu$ (without noise, $\ve = 0$);
{\bf (b)} for four noise levels $\ve$ (without drift, $\mu=0$).  Here,
one uses the minimal value of $\kappa_2$ over all lag times. }
\label{fig:VACF_ratio}
\end{figure}

For fBm without noise and drift, the ratio $\kappa_2/(2\kappa_1^2)$ is
minimal for $n = 1$ as expected.  Figure \ref{fig:VACF_ratio}
illustrates the behavior of this ratio as a function of $H$ (solid
blue curve).  One can observe the divergence at $H = 1/2$ as mentioned
earlier for Brownian motion.

In the presence of drift or noise, the ratio $\kappa_2/(2\kappa_1^2)$
is not necessarily minimal for $n = 1$.  As for TA MSD, we first find
the minimal value of $\kappa_2/(2\kappa_1^2)$ over all lag times and
then normalize it by the optimal ratio $\kappa_2^{\rm
opt}/(2\kappa_1^2)$.  The behavior of the resulting ratio
$\kappa_2/\kappa_2^{\rm opt}$ is shown on Fig. \ref{fig:VACF_ratio}
for different drift coefficients $\mu$ and noise levels $\ve$.  As
expected, there is no more divergence at $H = 1/2$ because the mean TA
VACF is not zero.  Similarly to TA MSD, the ratio is getting smaller
and weakly dependent on $H$ for large values of $\mu$ or $\ve$.
Moreover, the ratio becomes close to $1$ for large $\mu$.  This
behavior is expected from the very definition of the VACF as a measure
of correlations between displacements; in particular, the VACF is
constructed to access the drift.

\section*{Conclusion}

For a discrete non-centered Gaussian process $\x$, we studied the
problem of finding the symmetric matrix $\M_{\rm opt}$ that minimizes
a chosen cumulant moment $\kappa_m$ (e.g., the variance $\kappa_2$) of
the quadratic form $\chi = \frac12(\x^{\T} \M \x)$, under the
constraint of fixed mean value $\kappa_1$ of $\chi$.  The use of the
spectral representation (\ref{eq:kappa}) of the cumulant moments
$\kappa_m$ allowed us to reduce the original (possibly nonlinear)
optimization problem over the space of square matrices to a much
simpler optimization problem over the spectral parameters.  We gave
then an explicit solution of the reduced problem and constructed the
optimal matrix $\M_{\rm opt}$ in terms of the covariance matrix $\C$
and the mean vector $\x^0$ determining the Gaussian process.  At the
same time, this approach may be impractical for the inference of
unknown parameters from individual random trajectories because the
construction of the optimal form requires in general the complete
knowledge of the process.  Even if the optimal form can be
constructed, one still needs to interpret the outcome of such
measurement, for instance, to relate the mean value $\kappa_1$ of the
optimal form to the physical parameters of the process (such as
diffusion coefficient or drift).

In this light, the main practical result of the paper is the explicit
formula (\ref{eq:kappa_opt}) for the smallest achievable cumulant
moment $\kappa_m^{\rm opt}$.  This is the theoretical lower bound that
may serve as a quality benchmark in the optimality analysis of various
quadratic forms such as TA MSD, TA VACF, squared root mean square
displacement, power spectral density, etc.  In other words, the ratio
$\kappa_m/\kappa_1^m$ can be computed for a chosen quadratic form
(e.g., TA MSD) and a given class of Gaussian processes (e.g., fBm) and
then compared to the lower bound $\kappa_m^{\rm opt}/\kappa_1^m$.  The
difference may indicate whether the chosen quadratic form is well
adapted for the studied process.  A large difference would suggest
searching for other, more optimal, quadratic forms.

These optimality issues were illustrated for discrete fractional
Brownian motion altered by drift and independent Gaussian noise.  This
is a simple but rich model that incorporates anomalous features of the
dynamics (strong correlations between steps) and some measurement
imperfections such as electronic noises or cell mobility.  For this
model process, we computed the mean $\kappa_1$ and variance $\kappa_2$
of two quadratic forms broadly used for data analysis: TA MSD and TA
VACF.  The derived explicit formulas for $\kappa_1$ and $\kappa_2$
allowed us to analyze the influence of drift and noise onto
measurements.  We also compared the ratio $\kappa_2/\kappa_1^2$ to the
benchmark value $\kappa_2^{\rm opt}/\kappa_1^2$ of the optimal form.
In particular, we showed that the variance of the TA MSD exceeds the
optimal variance by at most $50\%$ for subdiffusive fBm ($H < 1/2$).
In turn, this variance increases dramatically for superdiffusive fBm
($H > 1/2$) suggesting that other quadratic forms may significantly
outperform the TA MSD in that case.

The spectral representation (\ref{eq:kappa}) allows one to tackle
other optimization problems.  In this paper, we focused on one
cumulant moment $\kappa_m$ of an even order $m$.  For odd cumulant
moments, the function $f$ in Eq. (\ref{eq:f}) is unbounded, and
supplementary constraints have to be added (e.g., one may restrict the
optimization problem to positive eigenvalues $\lambda_q$).  One may
also combine several constraints for simultaneous optimization of
different cumulant moments or their combinations (e.g., the skewness
$\kappa_3/\kappa_2^{3/2}$ or kurtosis $\kappa_4/\kappa_2^2$).
Finally, the spectral representation (\ref{eq:phi_k_log}) of the
characteristic function allows one to compute numerically the
probability density of a given quadratic form.

\section*{Acknowledgments}

Financial support from the ANR grant ``INADILIC'' is gratefully
acknowledged.


\appendix
\section{Multiple roots}
\label{sec:oneroot}

In this Appendix, we show that only one solution of
Eq. (\ref{eq:auxil1}) is compatible with Eq. (\ref{eq:lambda_q}).
From Eq. (\ref{eq:lambda_q}), one may express $c_q^2$ as
\begin{equation}
c_q^2 = \frac{\lambda^{m-1} - \lambda_q^{m-1}}{m\lambda_q^{m-1} - \lambda^{m-1}} > 0 ,
\end{equation}
that implies that both the numerator and denominator should be
positive: 
\begin{equation}
\label{eq:A_ineq}
\lambda_q^{m-1} < \lambda^{m-1} <  m\lambda_q^{m-1}.
\end{equation}
Since $m$ was assumed to be even, the second inequality implies that
$\lambda_q > 0$.  The substitution of $m\lambda_q^m$ from
Eq. (\ref{eq:auxil1}) into the above inequalities implies $\beta > 0$
and $\alpha (1-1/m)\lambda_q > \beta$, respectively.  One also
concludes that $\alpha > 0$.  Let us now consider the behavior of the
polynomial $g(x) = mx^m - \alpha x + \beta$ from
Eq. (\ref{eq:auxil1}).  One easily checks that this function admits a
single minimum on the positive semi-axis at $\lambda_c =
(\alpha/m^2)^{\frac{1}{m-1}}$.  Given that $g(0) = \beta >0$ and
$g(\infty) = +\infty$, positive roots of $g(x)$ exit if and only if
$g(\lambda_c) \leq 0$.  Moreover, when $g(\lambda_c) < 0$, there are
two distinct roots $\lambda_q^\pm$ such that $\lambda^- < \lambda_c <
\lambda_q^+$.  However, since $\lambda_c^{m-1} = \lambda^{m-1}/m$, the
inequality $\lambda^- <
\lambda_c $, written as $m [\lambda_q^-]^{m-1} < m \lambda_c^{m-1} =
\lambda^{m-1}$, contradicts to the second inequality in
(\ref{eq:A_ineq}).  As a consequence, only the solution $\lambda_q^+$
is compatible with Eq. (\ref{eq:lambda_q}).

\section{Variance of TA VACF}
\label{sec:VACF}

The coefficients of the variance of the TA VACF are
\begin{equation*}
\begin{split}
\tilde{c}_{\sigma,\sigma} & = 1 + \frac14 \biggl((n+1)^{2H} + (n-1)^{2H} - 2n^{2H}\biggr)^2 + R_H^{(2)} , \\
\tilde{c}_{\sigma,\ve}    & = 8(1-2^{2H-2}) + 4\frac{2^{2H-1}-1}{N-n-1} + \begin{cases} R_H^{(0)}  ~~ (n < N/2) ,\cr  0 \qquad (n\geq N/2) , \end{cases}  \\
\tilde{c}_{\ve,\ve}       & = 6 + \delta_{n,1} - \frac{2}{N-n-1} , \\
\tilde{c}_{\sigma,\mu}    & = 2 + (n+1)^{2H} + (n-1)^{2H} - 2n^{2H} + R_H^{(1)} , \\
\tilde{c}_{\mu,\ve}       & = \frac{4}{N-n-1}, \\
\end{split}
\end{equation*}
where
\begin{widetext}
\begin{equation*}
\begin{split}
R_H^{(0)}(n,N) & \equiv \biggl[4(2n+1)^{2H} + 4(2n-1)^{2H} - 6(2n)^{2H} - (2n+2)^{2H} - (2n-2)^{2H}\biggr] \\
& - \frac{4(2n+1)^{2H+1} + 4(2n-1)^{2H+1} - 6(2n)^{2H+1} - (2n+2)^{2H+1} - (2n-2)^{2H+1}}{2(N-n-1)} , \\
R_H^{(1)}(n,N) & \equiv \frac{1}{N-n-1}\sum\limits_{k=1}^{N-n-2} (N-n-1-k) \biggl(
2\bigl[(k+1)^{2H} + (k-1)^{2H} - 2k^{2H}\bigr] \\
& + (k+n+1)^{2H} + (k+n-1)^{2H} - 2(k+n)^{2H} + |k-n+1|^{2H} + |k-n-1|^{2H} - 2|k-n|^{2H} \biggr) , \\
%
\end{split}
\end{equation*}
\end{widetext}

\begin{widetext}
\begin{equation*}
\begin{split}
R_H^{(2)}(n,N) & \equiv \frac{1/2}{N-n-1}\sum\limits_{k=1}^{N-n-2} (N-n-1-k) \biggl\{ \biggl((k+1)^{2H} + (k-1)^{2H} - 2k^{2H}\biggr)^2 \\
& + \biggl((k+1+n)^{2H} + (k-1+n)^{2H} - 2(k+n)^{2H}\biggr)\biggl(|k+1-n|^{2H} + |k-1-n|^{2H} - 2|k-n|^{2H}\biggr) \biggr\}. \\
\end{split}
\end{equation*}
\end{widetext}

\end{document}